\documentclass[fleqn,12pt,twoside]{article}
\usepackage{espcrc1}

\usepackage{graphicx}
\usepackage[figuresright]{rotating}

\newcommand{\AmS}{{\protect\the\textfont2
  A\kern-.1667em\lower.5ex\hbox{M}\kern-.125emS}}

\title{Study of the ${}^{16}$O($\lowercase{p}$,$\gamma$) Reaction
      at Astrophysical Energies}

\author{C.~Barbieri\address[TRIUMF]{TRIUMF, 4004 Wesbrook Mall, Vancouver,
          British Columbia, Canada V6T 2A3}%
          \thanks{Email: \texttt{barbieri@triumf.ca} }
          \thanks{URL:   \texttt{http://www.triumf.ca/people/barbieri} }
       B.~K.~Jennings\addressmark[TRIUMF]
       }

\begin{document}

\maketitle

\begin{abstract}
 The Feshbach theory of the optical potential naturally leads to
a microscopic description of scattering
in terms of the many-body self-energy.
We consider a recent calculation of this quantity for ${}^{16}$O and
study the possibility of applying it at astrophysical
energies. The results obtained for the phase shifts
and the ${}^{16}$O($p$,$\gamma$) capture suggest that such studies are
feasible but the calculations require some improvement geared
to this specific task.
\end{abstract}

\section{Itroduction}

 The Feshbach theory of the optical potential provides a formal tool 
for developing a description of the scattering of nucleons from nuclei
in terms of the microscopic interaction.
 In its most standard application, this is done by 
reducing the Hilbert space to a subspace that contains
only the core nucleus and the additional particle in a scattering or 
bound single particle state.
Since the scattered particle is not allowed to occupy the orbitals filled
by  the core nucleons, this subspace can only exhaust the part of the 
one-body Hilbert space the lies above the Fermi energy).
 Several models of nucleon-nucleus scattering (such as cluster model or
folding potentials) implicitly use this approach and have been applied
with significant success.
However, the above limitation can be avoided by considering
a Fock space (without a definite number of particles) and applying 
the Feshbach formalism to a subspace that contains both the scattering
of a particle on top of the nuclear core and the possibility of
propagating a hole excitation~\cite{byron02}. The resulting optical potential
simply reduces to the usual many-body self-energy.
The properties of this particular choice of the optical model have been
reviewed by Mahaux and Sartor~\cite{Mahaux}.

The above considerations suggest that the technique of using
many-body Green's function
can make reasonable predictions for nucleon-nucleus scattering.
 However, most of its applications in nuclear physics have focused on the
the study of nuclear correlations~\cite{diba04}.
%
In this contribution, we consider the self-energy resulting from a recent
application of the self-consistent Green's function method the nucleus
of ${}^{16}$O~\cite{fadda,faddb} and explore its predictions for
proton-nucleus scattering.

\section{The model}

In Refs.~\cite{fadda,faddb}, the nuclear self-energy was computed 
within a model space $\cal{P}$ corresponding to the  harmonic oscillator
states of all orbitals up to the $pf$ shell plus the $1g_{9/2}$
orbital.  A parameter $b$=1.76~fm was employed and a
G-matrix interaction based on the Bonn-C potential~\cite{bonnc}
was used in the calculation.
 When expressed in coordinate space, this self-energy takes the form
\begin{equation}
  \Sigma^\star({\bf r},{\bf r}',\omega) =
     \sum_{\alpha, \beta \in \cal{P}} \; \phi_\alpha({\bf r})
       \left[
         \Sigma^{MF}_{\alpha \beta}(\omega) 
     +  \sum_p \frac{ \left( m^{p+}_\alpha \right)^* \; m^{p+}_\beta}
                      {\omega - \varepsilon^{p+}+ i\eta}
     +  \sum_h \frac{ m^{h-}_\alpha \; \left( m^{h-}_\beta \right)^*}
                      {\omega - \varepsilon^{h-} - i\eta}
       \right]
     \phi^*_\beta({\bf r}')  \; \;
\label{eq:Self-en}
\end{equation}
where $\phi_\alpha({\bf r})$ are the harmonic oscillator wave functions,
coupled to the nucleon spin with quantum numbers,
$\alpha=\{n_\alpha, l_\alpha, j_\alpha, m_\alpha\}$ and
the sum runs over all the orbits belonging to the model space~%
\footnote{The isospin degrees of freedom are not shown explicitly here.}.
%
%
We note that this choice for the space $\cal{P}$ is adequate
to study properties of the nuclear interior when considering low energy
(long range) excitations and accounting for
the spectral fragmentation~\cite{faddb}.
 However the gaussian like functions $\phi_\alpha({\bf r})$ are not
optimal when one is concerned with properties sensitive to the nuclear surface.

\begin{figure}
 \begin{center}
 \includegraphics[width=5in]{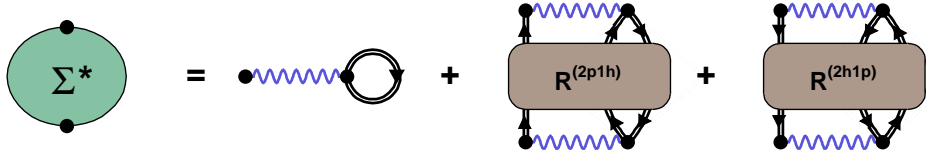}
 \end{center}
\vspace{-.5in}
\caption[]{\small
  Feynman diagram representation of the self energy. The first diagram 
 represent the Hartree-Fock like contribution to the mean field. The
 remaining ones describe core polarization effects in the particle
 ($R^{(2p1h)}$) and hole ($R^{(2h1p)}$) part of the spectrum.}
\label{fig:slef-en}
\end{figure}

In Eq.~(\ref{eq:Self-en}),  the mean field part
of the optical potential, $\Sigma^{MF}$, corresponds to the
Hartree-Fock diagram of Fig.~\ref{fig:slef-en}.
The remaining diagrams are usually referred as core
polarization contributions~\cite{Mahaux}. They were computed in
Ref.~\cite{faddb}  by including the coupling
of single particle motion (i.e. quasiparticles and quasiholes) to different
types of collective motion, namely excitations of the A-particle core,
two-particle and two-hole states.
%
%
At low enough energies 
they can be expressed as a discrete sum of poles, see
Eq.~(\ref{eq:Self-en}).
 There $\varepsilon^{i\pm}$
determine
the excitation energies of the resonances that do not have a mean field
character.

Thus, Eq.~(\ref{eq:Self-en}) gives us a model for the optical potential that 
acts in the full single particle Hilbert space.
However, before using it in practical applications
one has to correct for the fact that the calculations of Ref.~\cite{faddb}
did not include the electromagnetic interaction and were based on
a two-body realistic interaction, which is not sufficient to account
for spin-orbit splitting. 
 Hence, in this work we augment the 
self-energy~(\ref{eq:Self-en}) with the Coulomb potential for
a uniformly charged sphere of radius $R_c$~=~3.2~fm and add a correction
$U({\bf r},{\bf r}')$.
 The scattering equation takes a Schr\"odinger-like form~[$\hbar$=$c$=1 and $\mu$ is the reduced mass]
\begin{equation}
  \left\{
  \frac{- \nabla^2}{2 \mu} + V_{Coul.}({\bf r}) 
  \right\} \; \psi({\bf r})
  ~+~
  \int d{\bf r}' \; \left\{ 
       U({\bf r},{\bf r}') 
       ~+~ \Sigma^\star({\bf r},{\bf r}',E_{cm})
    \right\} \; \psi({\bf r}')
  ~=~ E_{cm} \; \psi({\bf r})   \; ,
\label{eq:Dyson}
\end{equation}
which, for $E_{cm} < 0$, also describes the bound states of ${}^{17}$F.
The potential
\begin{equation}
 U({\bf r},{\bf r}') = \sum_\alpha \; \delta\varepsilon_\alpha   \;
               \phi_\alpha({\bf r}')\phi^*_\alpha({\bf r}')  \; 
\label{eq:Urr1}
\end{equation}
is chosen in analogy to the work of Ref.~\cite{faddb} and acts by
simply shifting the energy of the principal mean field orbitals.
In practice, this corresponds to modifying the depth of the optical potential
independently for each different partial wave.
In the present work we found that one can fit the energy levels
of the orbitals in the $p$ and $sd$ shells by choosing
$\delta\varepsilon_{1p_{3/2}}$=-2.62~MeV,
$\delta\varepsilon_{1p_{1/2}}$=-2.25~MeV,
$\delta\varepsilon_{2s_{1/2}}$=+8.52~MeV,
$\delta\varepsilon_{1d_{5/2}}$=-3.75~MeV and 
$\delta\varepsilon_{1d_{3/2}}$=+8.4~MeV.

\section{Results}

\begin{figure}[tb]
 \begin{minipage}[t]{75mm}
 \includegraphics[width=2.7in]{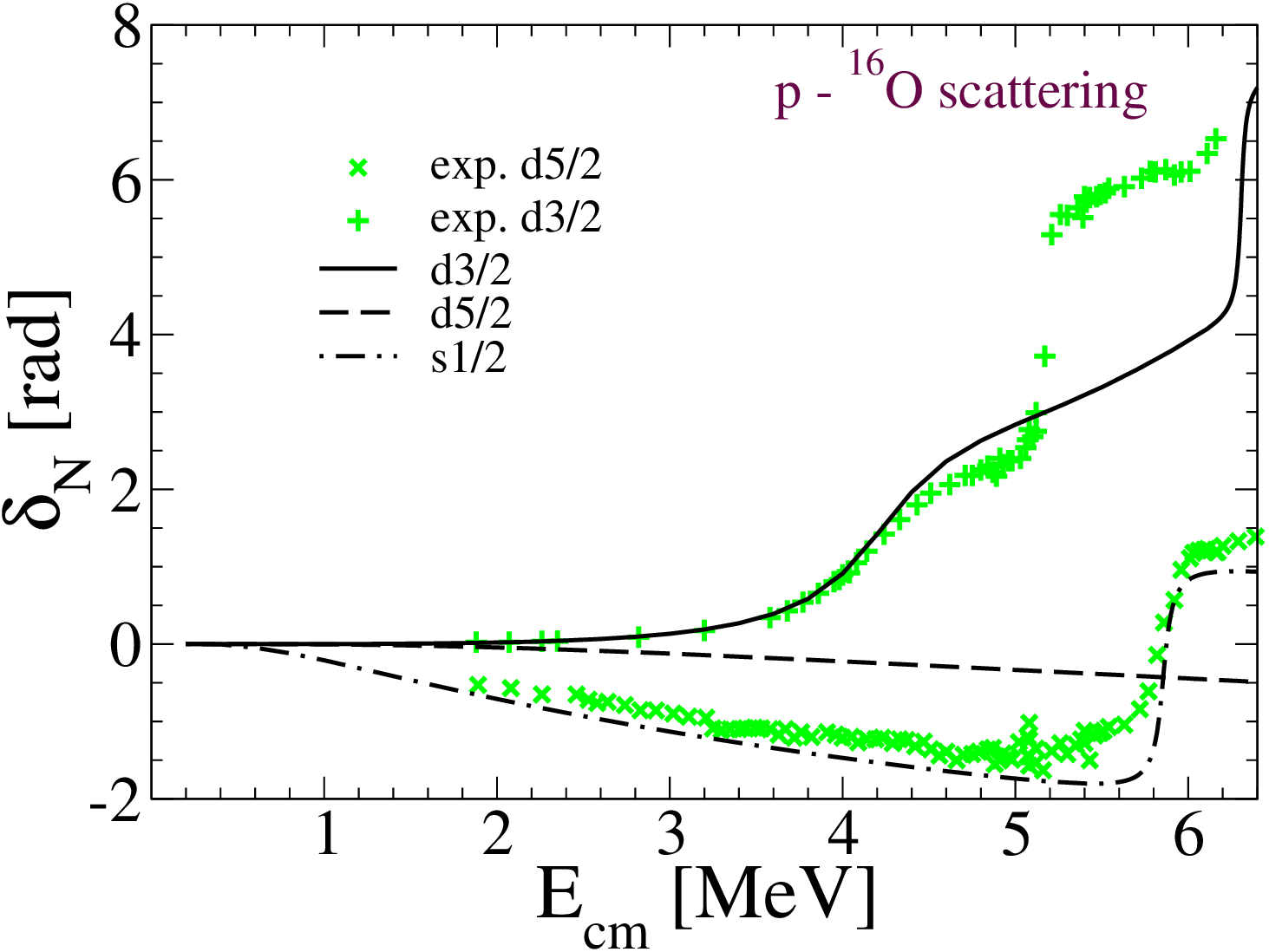}
 \caption{Phase shifts for proton~-~${}^{16}$O scattering in the 
 $s_{1/2}$, $d_{5/2}$ and $d_{3/2}$ partial waves,
 as a function of the center of mass energy.
 The experimental results are from Ref.~\cite{phase-sh}. }
 \label{fig:ph-sh}
 \end{minipage}
 \hspace{\fill}
 \begin{minipage}[t]{80mm}
 \includegraphics[width=2.7in]{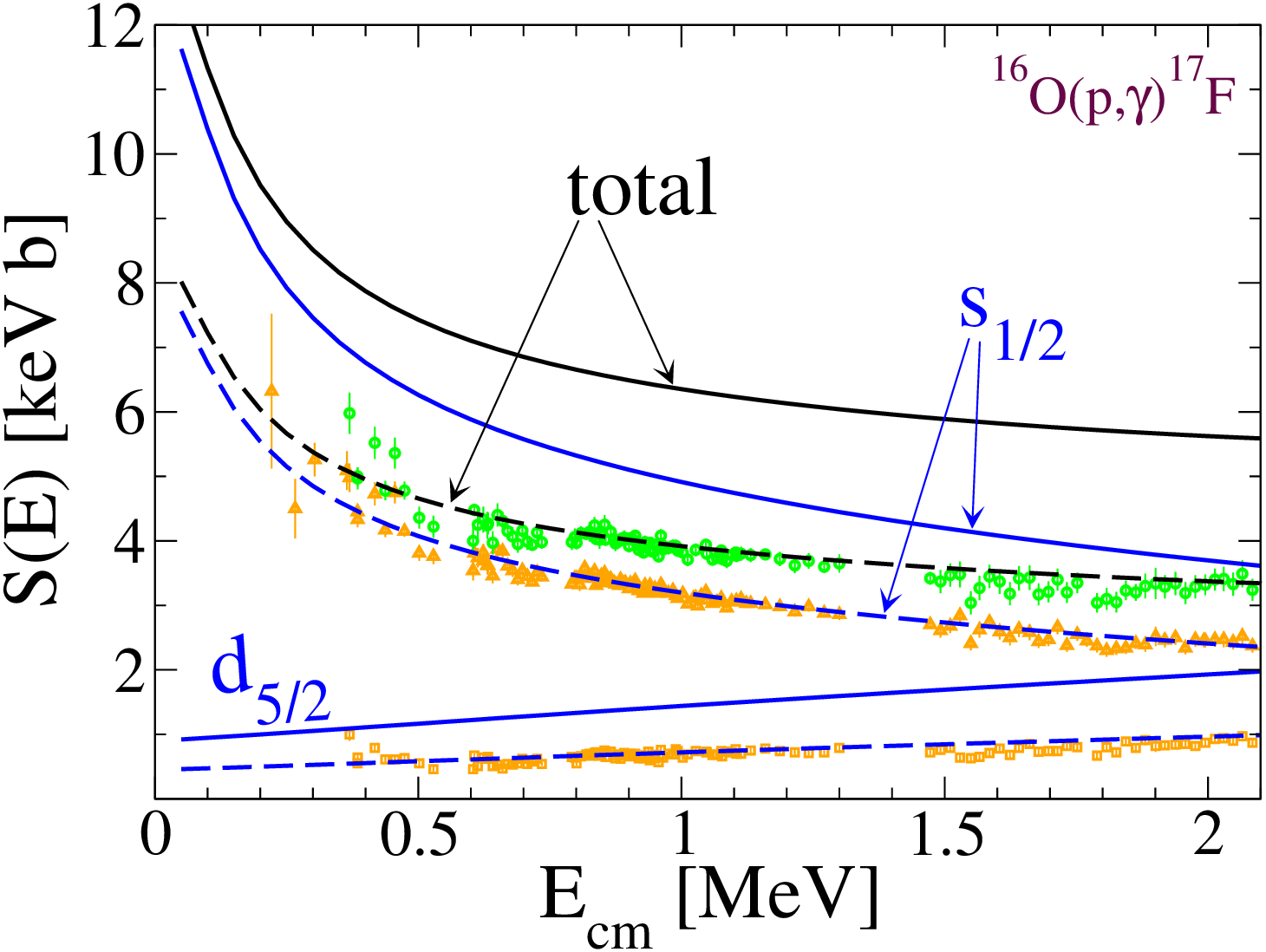}
 \caption{Astrophysical factor for the ${}^{16}$O($p$,$\gamma$) capture.
   The solid curves give the
  theoretical results for capture to the $d_{5/2}$ and $s_{1/2}$ states
  and the total capture.
   The dashed curves represent the same results after rescaling by a constant
  as described in the text.
   The experimental results are from Ref.~\cite{Morlock}.}
 \label{fig:S-factor}
 \end{minipage}
\end{figure}

  Figure~\ref{fig:ph-sh}  shows the phase shifts for the scattering
of positive parity waves resulting from Eq.~(\ref{eq:Dyson}).
The position of the bound states in ${}^{17}$F
and of the resonances shown in the plot has been forced to agree
with the experimental data employing the above choice of $U({\bf r},{\bf r}')$.
 This is necessary for obtaining the correct asymptotic behaviour of the
bound wave functions. The position of the first $s_{1/2}$ resonance
is determined by the first $\varepsilon^{p+}$ pole in Eq.~(\ref{eq:Self-en}).
This was originally predicted at 6.3~MeV, $\sim$0.5~MeV above the
experimental value.
 The background contribution to the phase shifts is instead a prediction of
the theory.  As it can be seen, the mean field tends to be slightly
too repulsive for the  $s_{1/2}$ partial wave and slightly too attractive
in the $d_{3/2}$ case. 
Considering the simple expansion of Eq.~(\ref{eq:Self-en}) in terms
of a few harmonic oscillator wave functions, the model
gives  a reasonable description of the phase shifts.
Eq.~(\ref{eq:Dyson}) was also solved for the valence orbitals of the
last proton bound  states of ${}^{17}$F. The
radial wave functions behave asymptotically as
\begin{equation}
  f_{lj}(r) \longrightarrow_{r \rightarrow \infty} 
       C_{lj}\frac{W_{-\eta, l+1/2}(r)}{r} \; ,
\label{eq:ANC_def}
\end{equation}
where $W_{-\eta, l+1/2}$ is a Whittaker function, $\eta$ the Sommerfield 
parameter and the asymptotic normalization constants (ANCs) are
predicted to be $C_{s_{1/2}}$=~98.2~fm$^{-1/2}$ and
$C_{d_{5/2}}$=~1.59~fm$^{-1/2}$.

The astrophysical factor for ${}^{16}$O($p$,$\gamma$) is computed
from these wave functions and the solutions of Eq.~(\ref{eq:Dyson})
for the relevant scattering waves.
The results for the transitions
to the bound states of ${}^{17}$F
are reported in Fig.~\ref{fig:S-factor} with solid lines.
 The curves
overestimate the experimental results, due to the large values
obtained for the ANCs. 
 In contrast, the relative spectroscopic factors are found to be
$0.921$ for the $s_{1/2}$ state and $0.878$ for the $d_{5/2}$ case.
This
agrees with what is expected from the halo nature of these orbitals.
Moreover, we find that the shape of the theoretical curves agree well
with the experimental data if we rescale the result for
capture to the $s_{1/2}$ state by $0.63$ and the one
for $d_{5/2}$ by $0.49$. This
is depicted by the dashed lines in Fig.~\ref{fig:S-factor}.

At zero energy, the astrophysical factor is determined by ANCs.
 The analytical study of the asymptotic wave functions done in
Ref.~\cite{BayeDes} yields the following approximation,
\begin{equation}
  S(0) ~=~ 0.37 C_{s_{1/2}}^2  ~+~ 1.58 \times 10^{-3} C_{d_{5/2}}^2 \; .
\label{eq:S0res}
\end{equation}
After rescaling $C_{s_{1/2}}$ and $C_{d_{5/2}}$ as described above
Eq.~(\ref{eq:S0res}) gives $S(0)$=10.37~keV~b,
in agreement with their estimate~\cite{BayeDes}.

The discrepancies found between theory and experiment are not unexpected.
Part of these are due to the fact that the expansion of Eq.~(\ref{eq:Self-en}),
originally chosen to study properties sensitive to the interior of the
nucleus, gives a poor description of the nuclear surface to which halo
states are particularly sensitive.
 A better choice of the basis would involve repeating completely the
calculations of Ref.~\cite{faddb}. Nevertheless, it is plausible that
this approach would lead to useful predictions of the scattering process.

\hspace{.5cm}

We would like to acknowledge useful discussions with J.-M.~Sparenberg.
This work was supported by the Natural
Sciences and Engineering Research Council of Canada (NSERC).

\end{document}